\documentclass{jetpl}
\twocolumn
\title{Spin-motive  force and orbital-motive force:
 from magnon BEC to chiral Weyl superfluids}

\rtitle{Spin-motive  force and orbital-motive force}

\sodtitle{Spin-motive  force and orbital-motive force}

\author
{G.E. Volovik $^{+\#}$ \/\thanks{e-mail: volovik@boojum.hut.fi}
}

\rauthor{
G.E.Volovik
}

\sodauthor{
Volovik
}

\address{
$^{+}$ Low Temperature Laboratory, Aalto University, School of Science and
Technology, P.O. Box 15100, FI-00076 AALTO, Finland
\\
$^{\#}$ Landau Institute for Theoretical Physics RAS, Kosygina 2, 119334 Moscow, Russia
}

\abstract{ Spin-motive geometric force acting on electrons in metallic ferromagnets is extended to spin-motive force in magnon BEC, which is represented by phase-coherent precession of magnetization,  and to the orbital-motive force in superfluid $^3$He-A. In $^3$He-A  there are two contributions to the orbital-motive force. One of them comes  from the
chiral nature of this liquid. Another one originates from chiral Weyl fermions living in the vicinity of the topologically protected Weyl points, and is related to the phenomenon of chiral anomaly.}

\begin{document}

\maketitle

 \section{Introduction}

A spin-motive force is a  force acting on electron from the magnetization.
It was introduced for metallic ferromagnets 
\cite{Berger1986,Volovik1987,Barnes2007,Yang2009,Yamane2011,Hayashi2012,Nakabayashi2013}, 
where it
reflects the conversion of the magnetic energy of a ferromagnet
into the electrical energy of the conduction electrons. The spin-motive force includes the geometric force which originates from the Berry phase in spin dynamics \cite{Volovik1987}.
In metals, the spin-motive force reflects the intricate coupling between the electric current and the collective magnetic degrees of freedom. The inverse phenomenon is the effect of electric currents on the dynamics of magnetization in ferromagnets.  It is now a concept relevant to electronic devices.

The collective magnetic degrees of freedom exist not only in ferromagnets, but also in other systems, such as
Bose-Einstein condensation of magnons manifested in the phase-coherent precession (the so-called magnon BEC, see review \cite{BunkovVolovik2013}) and the chiral Weyl superfluid $^3$He-A, which is the orbital ferromagnet. We extend here 
the concept of the spin-motive geometric force to spin-motive force in magnon BEC and to the geometric orbital-motive force in $^3$He-A. In $^3$He-A  there are two contributions to the orbital geometric force: one is related to the
chiral nature of this liquid and another one is the property of the topological Weyl superfluid, which experiences the chiral anomaly.

\section{Spin-motive force in magnon BEC}

 The geometric force acting on atoms of $^3$He in the precessing state can be obtained from the geometric spin-motive force discussed in Ref. \cite{Volovik1987} for ferromagnets:
\begin{equation}
F^{\rm spin-motive}_i=\frac{\hbar}{2} (n_+ - n_-)\hat{\bf m}\cdot(\partial_t \hat{\bf m}\times \nabla_i\hat{\bf m}) \,.
\label{SMF1}
\end{equation}
Here ${\bf F}^{\rm spin-motive}$ is the geometric spin-motive force; $\hat{\bf m}$ is the unit vector parallel to the local spin magnetization; $n_+ - n_-$ is spin polarization.
The origin of this force is the Berry phase.The collective spin dynamics is governed by $SU(2)$ symmetry group. The conduction electrons feel the $U(1)$ subgroup of spin rotations about the local magnetization as the gauge field. This is one of the numerous examples of emergent gauge fields in many-body systems
(other types of the gauge field originating from spin degrees of freedom can be found in 
\cite{DzyaloshinskiiVolovik1978,MineevVolovik1992}). The `electric' field component of the effective $U(1)$ gauge field is
\begin{equation}
E^{\rm geom}_i=\frac{\hbar}{2} \hat{\bf m}\cdot(\partial_t \hat{\bf m}\times \nabla_i\hat{\bf m}) \,.
\label{Efield}
\end{equation}
 This electric field drives spin-up and spin-down conduction electrons in opposite directions, inducing a spin current. In the presence of spin polarization due spontaneous magnetization emerging in metallic ferromagnets, this field produces the spin-motive force  in Eq.(\ref{SMF1}) giving rise to electric current. 

 Eq.(\ref{SMF1}) can be also applied to magnon BEC -- the spontaneously precessing magnetization, which emerges in supefluid  $^3$He-B in pulsed NMR experiments.  Shortly after the pulse the state of the coherent spin
precession is formed: 
\begin{eqnarray}
{\bf S}({\bf r},t)=S\hat{\bf m}({\bf r},t)
\,,
\\
\hat{\bf m} ({\bf r},t)=\hat{\bf z}\cos\beta({\bf r}) +\sin \beta({\bf r})  (\hat{\bf x}\cos\omega t +
 \hat{\bf y}\sin\omega t ) \,.
\label{precession}
\end{eqnarray}
Here $\omega$ is the global frequency of free precession: it is the same in the whole sample even if the tipping angle of the deflected magnetization $\beta({\bf r})$ is coordinate dependent. The phase coherent precession is the signature of magnon BEC. The direction of magnetic field -- the axis of precession -- is assumed along the axis $\hat z$. The modulus of the deflected spin is the same as in equilibrium non-precessing states: $S=\chi H/\gamma$, where $\chi$ is magnetic susceptibility, and $\gamma$ is the gyromagnetic ratio of $^3$He atom. Since $S$ is the difference between the density of spin-up and spin-down atoms in applied magnetic field
$S=\frac{\hbar}{2} (n_+ - n_-)$, one obtains from Eq.(\ref{SMF1}) the following spin-motive force -- the transfer of linear momentum density from the collective magnetic subsystem (magnon BEC)  to the mass current:
\begin{equation}
F^{\rm spin-motive}_i=  S\hat{\bf m}\cdot(\partial_t \hat{\bf m}\times \nabla_i\hat{\bf m}) \,,
\label{SMF2}
\end{equation}
which for the space coherent precession in Eq.(\ref{precession}) reads:
\begin{equation}
F^{\rm spin-motive}_i=S\hat{\bf m}\cdot(\partial_t \hat{\bf m}\times \nabla_i\hat{\bf m})=S\omega \nabla_i \cos\beta=-\mu \nabla_i n\,.
\label{SMFprecession}
\end{equation}
Here in the last equation we used the language of magnon BEC, in which $n=(S-S_z)/\hbar$ is the magnon density and $\mu=\hbar\omega$ plays  the role of their chemical potential. In a full equilibrium, i.e. 
without precession, the chemical potential of magnons is zero, since their number is not conserved. In   $^3$He-B the life time of coherent precession is large compared to the time of the formation of the coherent state. That is why for such quasiequilibrium state the magnon number is quasi-conserved. 

In the limit of vanishing dissipation the magnon chemical potential  $\mu$ is nonzero and is well determined, being equal to the global frequency of precession, which in turn depends on the number of the pumped magnons. 
In this limit, when losses of magnetization can be neglected and the spin projection $S_z$ on magnetic field can be considered as conserved quantity, the precessing state becomes an example of the system with spontaneously broken time translation symmetry and off-diagonal long range order (ODLRO). In a given case the ODLRO is based on the operator of spin creation $\left<S_+\right> =S\sin \beta   ~ e^{i\mu t+i\alpha}$ 
\cite{Wilczek2013}. In principle, all the systems
with off-diagonal long range order demonstrate the  broken time translation symmetry in the ground state, assuming that it is the ground state under condition of fixed value of some conserved (or quasi-conserved) quantity, or the fixed value of the corresponding chemical potential. 

Example of ODLRO is also provided by the precession of partially trapped vortex line  in $^3$He-B, where hours long oscillations have been experimentally observed
\cite{Zieve1992,MisirpashaevVolovik1992,Packard1998}. In the limit of vanishing dissipation the projection $L_z$ of the orbital angular momentum of the liquid can be considered as  conserved quantity. Then the precessing state of a vortex
can be represented in terms of the ODLRO based on the operator of creation of the orbital angular momentum
$\left<L_+\right> \propto  e^{i\omega t+i\alpha}$. This coherent precession can be also considered as BEC of excitations propagating along the vortex -- Kelvin waves or kelvons.

\section{Orbital-motive force and chiral anomaly}

The Berry-phase geometric force emerges not only from spin. It may also come from other internal degrees of freedom, such as pseudospin -- valley spin, isotopic band spin, Bogoliubov-Nambu spin, etc. 
The low-frequency dynamics of the order parameter in the broken symmetry systems often induces the transport of mass, electric charge and spin, see e.g. \cite{Tserkovnyak2005}.
We consider here the geometric force generated by dynamics of the orbital angular momentum. Example is provided by chiral superfluid  $^3$He-A, which is polarized in terms of the orbital angular momentum (the chirality of  $^3$He-A is observed in recent experiments, see
\cite{WalmsleyGolov2012,VolovikKrusius2012,Kono2013}). The corresponding geometric effect of transformation of orbital degees of freedom to the mass current is known under the name of Mermin-Ho relation \cite{Mermin-Ho}, which leads to the following dynamic equation for superfluid velocity \cite{VollhardtWolfle}:
\begin{equation}
 \partial_t(m{\bf v}_{\rm s}) -   \nabla\mu=
      \frac{\hbar}{2}~ \hat{\bf l}\cdot(\partial_t \hat{\bf l}\times \nabla\hat{\bf l})
\,.
\label{Mermin-HoEqDyn}
\end{equation}
Here ${\bf v}_{\rm s}$ is superfluid velocity and $\hat{\bf l}$ is the unit vector parallel to the local orbital momentum density ${\bf L}$. The inverse effect -- oscillations of the orbital vector $\hat {\bf l}$ generated by applied heat current -- have been observed in \cite{Paulson1976} (see also \cite{Volovik1978}).

At $T=0$ the orbital momentum of superfluid $^3$He-A is fully polarized with momentum $\hbar/2$ per each atom
(or $\hbar$ per each Cooper pair). This means that $n_+ -n_-=n_3$, where $n_3$ is the particle density of $^3$He atoms, and ${\bf L}=(\hbar/2) n_3\hat{\bf l}$.
When multiplied by $n_3$, the right-hand side of 
Eq.(\ref{Mermin-HoEqDyn}) gives the orbital-motive geometric force acting from the collective
orbital momentum degrees of freedom to the mass current:
\begin{equation}
F^{\rm orbital-motive}_i= \frac{\hbar}{2} n_3 \hat{\bf l}\cdot(\partial_t \hat{\bf l}\times \nabla_i\hat{\bf l})
\label{OrbitalMotiveForce}
\end{equation}

However, this is not the whole story. Chiral superfluid  $^3$He-A belongs to the class of topological Weyl superfluids. These substances have topologically protected gap nodes in the fermionic spectrum -- Weyl points, which represent the Berry phase magnetic monopoles in momentum space \cite{Volovik1987b}. Close to any of two Weyl points, fermionic quasiparticles behave as chiral Weyl fermions. As a consequence, there is another geometric orbital-motive force, which comes from the Adler-Bell-Jackiw chiral anomaly \cite{Volovik2003} (on anomalies in dynamics of chiral  liquids including the dense quark-gluon matter in QCD and hypothetical Weyl materials see 
Refs. \cite{Basar2013,BasarYee2013,Son2012,Son2013,VivekAji2012}
 and references therein):
\begin{equation}
F^{\rm chiral-anomaly}_i = \frac{ 1}{2\pi^2} p_F \hat l_i {\bf B} \cdot {\bf E} = - \frac{p_F^3}{2\pi^2 \hbar^2}\hat l_i
\left(\partial_t\hat{\bf l}\cdot(\nabla\times \hat{\bf l})\right)
\,.
\label{MomentumProductionGeneral}
\end{equation}
Here ${\bf B}=(p_F/\hbar)\nabla\times\hat{\bf l}$ and   ${\bf E}= (p_F/\hbar)
\partial_t
\hat{\bf l}$ are effective `magnetic'  and `electric' fields acting on
chiral Weyl quasiparticles in $^3$He-A. This is another example of emergent $U(1)$ gauge field in condensed matter. In a given case the gauge field emerges as collective mode of the fermionic vacuum, which contains the Berry magnetic monopoles in ${\bf p}$-space. 

Though the two geometric forces have different origin, they have common property, which is manifested in their effect on dynamics
of a continuous vortex  -- doubly quantized vortex, which represents the skyrmion in the $\hat{\bf l}$-field.
When integrated over the smooth core of the vortex-skyrmion moving with velocity $ {\bf v}_{\rm L}$, these two forces give similar contributions to
the non-dissipative Magnus-like force acting on the vortex-skyrmion in direction perpendicular to its velocity:
\begin{equation}
{\bf F}_{\rm transverse}=   2\pi \hbar 
 \left(n_3 -  \frac{p_F^3}{3\pi^2 \hbar^3} \right) {\hat {\bf z}}  \times {\bf v}_{\rm L} \,.
\label{MagnusForce}
\end{equation}
Here axis $\hat {\bf z}$ is along the vortex line, and the force is per unit length of the vortex.

The first term (with prefactor $n_3$) is similar to the Berry-phase Magnus force acting on magnetic vortices in magnets, see \cite{Thiele1973,NikiforovSonin1983,Volovik1986}.

The second term, which comes from the chiral anomaly, is known under the name of spectral flow force or Kopnin force.
In the weak coupling regime the density of the liquid in superfluid state only slightly deviates from its value in the normal state, $|n_3 - p_F^3 / 3\pi^2 \hbar^3| \ll n_3$. As a result the two terms in Eq.(\ref{MagnusForce}) almost exactly cancel each other. This cancellation is observed in Manchester experiment on $^3$He-A \cite{BevanNature}, which demonstrates the important role of the chiral anomaly for the vortex dynamics
in chiral superfluds.

With increasing interaction the difference between two contributions increases. Finally one reaches the topological quantum phase transition, at which the Weyl pont nodes merge and annihilate each other, and fermionic spectrum becomes fully gapped. After that the chiral anomaly contribution to the Magnus force is nullified leaving only the $n_3$-contribution in Eq.(\ref{MagnusForce}).
The interplay of the two types of chirality  -- chirality of the whole liquid (the chiral `vacuum') and chirality of its Weyl fermionic quasiparticles (which play the role of `matter' in the chiral vacuum) --  leads to a number of `paradoxes', including that of the orbital angular momentum of $^3$He-A (see Ref. \cite{Sauls2011} and references therein, \cite{Volovik1995}).

Two competing effects also govern the dynamics of singular vortices in fermionic systems (superfluid $^3$He-B and superconductors). The chirality there is produced by the vortex itself, which violates time reversal symmetry, while the spectral flow is carried by fermion zero modes localized within the vortex core \cite{Volovik2003}.

\section{Discussion}

Spin-motive geometric force transfers the momentum from  the collective magnetization modes to the conduction electrons in ferromagnetic metals and to the atoms of $^3$He in the state of magnon BEC.
In the latter case it is expressed via the density of magnons and their chemical potential in Eq.(\ref{SMFprecession}).
The mass current induced by this geometric force may be responsible for some effects observed in dynamics of magnon BEC
 \cite{Heikkinen2013,Eltsov2013}.

The analog of the spin-motive force in chiral superfluid is the orbital-motive force.
In superfluid $^3$He-A, there are two type of the orbital-motive forces. One of them comes  from the
chirality of the liquid. The second one comes from the chiral fermionic quasiparticles (Weyl fermions) living near the topologically protected Weyl points; it is the consequence of the phenomenon of chiral anomaly. These two forces results in the non-dissipative transverse forces acting on vortex-skyrmion in Eq.(\ref{MagnusForce}): Magnus force and spectral-flow force correspondingly. These two forces nearly compensate each other in the weak coupling regime. At the topological quantum phase transition, when the Weyl points annihilate each other and chiral liquid becomes fully gapped, the spectral flow force disappears.

The interplay of the orbital-motive and spin motive geometric forces is expected for magnon BEC state in $^3$He-A and especially for the A$_1$-phase of superfluid $^3$He, which is simultaneously spin polarized chiral superfluid, orbital chiral liquid and topological Weyl superfluid with chiral fermions.

\section*{\hspace*{-4.5mm}ACKNOWLEDGMENTS}
I acknowledge financial
support  by the EU 7th Framework Programme
(FP7/2007-2013, grant $\#$228464 Microkelvin) and by the Academy of
Finland through its LTQ CoE grant (project $\#$250280).


\end{document}